\documentclass[3p,times]{elsarticle}

\usepackage{ecrc}

\usepackage{epstopdf}

\usepackage{bbm}
\usepackage{amssymb}
\usepackage{ifpdf}
\usepackage{color}
\usepackage{graphicx,epsfig}

\newcommand{\Z}{{\mathbb{Z}}}

\volume{00}

\firstpage{1}

\journalname{Nuclear Physics A}

\runauth{Uwe-Jens Wiese}

\jid{nupha}

\jnltitlelogo{Nuclear Physics A}

\usepackage{graphicx}
\usepackage{amsmath,amssymb}

\begin{document}

\begin{frontmatter}



\title{Towards Quantum Simulating QCD}

\author{Uwe-Jens Wiese}

\address{Albert Einstein Center for Fundamental Physics, Institute for 
Theoretical Physics, Bern University, Sidlerstrasse 5, 3012 Bern, Switzerland}

\begin{abstract}
Quantum link models provide an alternative non-perturbative formulation of
Abelian and non-Abelian lattice gauge theories. They are ideally suited for
quantum simulation, for example, using ultracold atoms in an optical lattice.
This holds the promise to address currently unsolvable problems, such as the
real-time and high-density dynamics of strongly interacting matter, first in
toy-model gauge theories, and ultimately in QCD.
\end{abstract}


\end{frontmatter}

\section{Introduction}

Strongly interacting quantum matter may undergo intriguing real-time evolution.
Prominent examples range from the expansion of an ultracold atomic gas released
from a trap, or the out-of-equilibrium dynamics of a strongly correlated 
electron system, to an expanding quark-gluon plasma produced in a heavy-ion
collision. Due to the enormous dimension of the Hilbert space, which increases
exponentially with the system size, understanding the dynamics of large strongly
coupled quantum systems is a notoriously hard problem. Even when considered in
thermal equilibrium, Euclidean time Monte Carlo simulations of such systems 
often suffer from sign problems. For example, the fermionic Hubbard model away 
from half-filling or QCD at high quark density are currently inaccessible 
to first principles Monte Carlo simulations due to very severe fermion sign 
problems, thus preventing a better understanding of high-temperature 
superconductors or the dense cores of neutron stars. Real-time simulations of 
strongly interacting systems are also currently beyond reach, due to very 
severe complex weight problems in the corresponding real-time path integrals.
Some sign problems even fall in the complexity class of NP-complete problems
\cite{Tro05}, which can be solved in polynomial time on a hypothetical 
``non-deterministic'' computer, but for which no polynomial-time algorithm is 
known on an ordinary classical computer modeled by a Turing machine. Since one
expects that NP $\neq$ P (where P is the complexity class of problems that are
solvable in polynomial time on a classical computer), a generic solution for
general sign problems is unlikely to exist. This does not exclude that specific
sign problems may indeed be solvable on classical computers. In fact, several
severe sign or complex action problems have been solved with the meron-cluster 
algorithm \cite{Bie95,Cha99,Alf01} or with the fermion bag approach 
\cite{Cha10,Cha12,Huf14}. Even the real-time evolution of a large strongly 
coupled quantum spin system, whose dynamics is entirely driven by measurements,
has recently been simulated successfully with a cluster algorithm \cite{Ban14}.

While it is not excluded that the doped Hubbard model or QCD at high quark 
density can be simulated on classical computers, it is already clear that a 
universal quantum computer could indeed overcome several of the limitations of 
classical computers \cite{Llo96,Sho97}. In particular, since it operates with 
quantum hardware, a quantum computer can naturally manipulate complex amplitudes
and thus does not suffer from sign or complex weight problems. Although it is 
not known whether a quantum computer could solve NP-hard problems, it would be 
extremely useful for deepening our understanding of the dynamics of strongly 
coupled quantum systems. While theoretical constructions for quantum computers 
based on ultracold trapped ions exist already for some time \cite{Cir95}, it is 
difficult to predict when powerful quantum computers may become available. The 
D-wave devices based on a network of superconducting flux qubits have been used 
to operate the quantum adiabatic algorithm \cite{Far00} on random instances of 
an Ising spin glass \cite{Boi13}. A comparison with simulated classical and 
quantum annealing algorithms led to the conclusion that the D-wave machines 
indeed perform quantum rather than classical annealing, but are not yet 
competitive with classical computers.

As early as 1982, Feynman proposed to use specifically designed quantum devices 
to simulate other quantum systems \cite{Fey82}. Based on the experimental
breakthrough of realizing ultracold Bose-Einstein condensates 
\cite{And95,Dav95}, Feynman's vision of quantum simulators is becoming a 
reality. Although they are not universal quantum computers, but special purpose
devices that are designed to mimic a specific quantum system, quantum simulators
have the potential to drastically improve our understanding of strongly coupled
quantum systems \cite{Cir12}. Quantum simulators have been constructed using 
ultracold atoms in optical lattices \cite{Lew12,Blo12}, trapped ions 
\cite{Bla12}, photons \cite{Asp12}, or superconducting circuits on a chip 
\cite{Hou12}. One distinguishes digital \cite{Llo96} and analog 
\cite{Jak98} quantum simulators. A digital quantum simulator is a controllable 
many-body system, which encodes the state of the simulated system as quantum
information, and executes a programmable sequence of quantum gate operations, 
following a stroboscopic Trotter decomposition, in order to represent its 
dynamics. Analog quantum simulators, on the other hand, realize continuous
real-time evolution. They are limited to simpler interactions, but are more 
easily scalable to large system size. It was an experimental breakthrough when
ultracold atoms in an optical lattice were used as an analog quantum simulator 
to address the quantum phase transition that separates a Mott insulator state 
(with localized particles) from a superfluid in the bosonic Hubbard model 
\cite{Gre02}. The optical lattice is realized with counter-propagating laser 
beams, whose intensity determines the amplitude for hopping between neighboring 
lattice sites. The quantum simulation has been validated in comparison with 
accurate quantum Monte Carlo simulations \cite{Tro10}, which are possible 
because the Bose-Hubbard model does not suffer from a sign problem. The next 
challenge in this field is to quantum simulate the fermionic Hubbard model, in 
order to decide whether it can describe high-temperature superconductivity. 
This requires reaching even lower temperatures in the quantum simulator than 
the ones achieved until now.

When we use a quantum simulator to learn more about the dynamics of a strongly
coupled system, we are actually doing an experiment. As usual in quantum 
physics, the device is first prepared in an initial state, it then evolves in
time driven by its Hamiltonian, and finally it is analyzed by measurements. By 
repeating the quantum simulation experiment many times, and by averaging over 
the measurement results, one can compare it with the predictions of quantum 
mechanics. Successful quantum simulations benefit from close collaboration 
between theorists and experimentalists, from appropriately designing to 
accurately performing and theoretically validating the experiments. Since 
quantum simulators are currently far from being standard devices, a theorist 
will want to know whether they are an appropriate tool for theoretical physics. 
Ultimately, this will depend on their accuracy and reliability, which is 
currently limited but should improve significantly with time. Certainly, 
classical computers have matured to the point where a theorist performing, for
example, a Monte Carlo simulation does not necessarily think of herself as an 
experimentalist, despite the fact that she is operating a very sophisticated 
instrument. Instead, she may think of a classical computer as a ``black box''
that perfectly realizes the computational model of a Turing machine, which just
extends our mathematical capabilities. It is currently unclear when (or even
whether) quantum simulators or quantum computers will reach a similar status,
but there is no reason to be overly pessimistic. Even our brain, which is likely
the most sophisticated piece of ``hardware'' in the universe, is not always
completely reliable, but a theorist is not reluctant to use it as best as
possible. Quantum simulators and ultimately quantum computers promise to 
significantly extend our brain's capabilities in qualitatively new directions,
and we should not be reluctant to push their development forward as much as
possible.

Proposals for quantum simulator constructions already exist for some simple 
bosonic \cite{Ret05,Hor10,Jor12} and fermionic \cite{Cir10,Ber10,Boa11,Maz12} 
field theories. Hence it is natural to ask whether our understanding of strongly
coupled systems in nuclear and particle physics may benefit from quantum 
simulation. This certainly
is a long-term project, which can only be realized in small steps. Before one 
addresses full QCD, one should gain experience with simpler toy-model gauge 
theories. Such models may exist in a lower space-time dimension, they may 
have a simpler $U(1)$ or $SU(2)$ gauge group, instead of QCD's $SU(3)$ gauge 
symmetry, or they may exist on a lattice away from the continuum limit. They 
may also have a reduced $\Z(2)$ instead of Nature's $SU(2)_L \times SU(2)_R$ 
chiral symmetry, or they may have bosonic instead of fermionic baryons. All 
this implies that, for quite some time, we should not expect quantum simulations
to yield quantitative results that are directly relevant to nuclear or particle 
physics. Still, the toy-models mentioned above share different qualitative 
features with QCD, and will have similar dynamics, at least in some respects. In
particular, like QCD, simple toy-model gauge theories may have a spontaneously 
broken chiral symmetry, which is restored at finite baryon density. They may 
also display color superconductivity at high baryon density, with or without 
color-flavor locking. One can also imagine that a gauge theory quantum 
simulator can mimic qualitative features of heavy-ion collisions. Since we have
no other way of reliably investigating QCD's real-time dynamics from first 
principles, the construction of quantum simulators for Abelian and non-Abelian 
gauge theories, with or without fermionic matter, is timely and most promising 
\cite{Bue05,Wei10,Kap11,Zoh11,Szi11,Tag12,Zoh12,Ban12,Ban13,Zoh13,Tag13,Zoh13a,Zoh13b,Mar13,Mar14,Wie13}. Once such systems are realized in the laboratory, 
they will also become interesting objects of study in their own right.

If quantum simulations were ultimately limited to toy-model gauge theories,
they would be of limited interest for nuclear and particle physics. While the
first steps towards gauge theory quantum simulations will be taken in simpler
gauge theories, quantum link models provide a long-term vision for how to
eventually realize quantum simulations of QCD. The first $U(1)$ and $SU(2)$ 
gauge theories formulated with (generalized) quantum spins were constructed by 
Horn \cite{Hor81}, and further investigated by Orland and Rohrlich under the 
name of gauge magnets \cite{Orl90}. Quantum link models \cite{Cha97,Bro99,Bro04}
provide an alternative to Wilson's lattice gauge theory \cite{Wil74,Kog79,Kog83}
for defining QCD beyond perturbation theory. Both in the Wilson formulation and 
in quantum link models, the basic gauge degrees of freedom are associated with 
the links connecting nearest-neighbor sites of a lattice that serves as a 
regulator of ultra-violet divergences. In Wilson's lattice gauge theory, the 
link variables are parallel transporter matrices that take values in the gauge 
group. In particular, in lattice QCD they are $SU(3)$ matrices associated with 
each link. Since the $SU(3)$ group space is a continuous manifold, in the 
Wilson theory the local link Hilbert space is infinite-dimensional. This poses 
a severe challenge for quantum simulations with ultracold atoms, because in 
practice only a small number of individual atomic states 
can be used to encode quantum information. In contrast to Wilson's lattice 
gauge theory, quantum link models are ideally suited for quantum simulation 
because they have a finite-dimensional Hilbert space per link. While $SU(N)$ 
quantum links are still $N \times N$ matrices, unlike in the Wilson theory, 
their matrix elements are non-commuting operators, which are generators of an 
$SU(2N)$ embedding algebra. In the quantum link formulation of QCD, the gauge 
degrees of freedom are described by a 20-dimensional representation of an 
$SU(6)$ embedding algebra. Hence five qubits (with a $2^5 = 32$-dimensional 
Hilbert space) are more than one needs to represent the quantum information 
encoded in an $SU(3)$ quantum link, while a Wilson-type link variable exists 
in an infinite-dimensional Hilbert space. Thanks to universality, QCD
can be regularized in several different ways, and it is useful to choose the
most appropriate regularization, depending on the question one wants to address.
Certainly, when one wants to study QCD at very high energy, thanks to asymptotic
freedom, one can use perturbation theory, which is best handled using 
dimensional regularization and renormalization applied to the standard QCD
Lagrangian. On the other hand, when one wants to address non-perturbative 
questions concerning, for example, static properties of hadrons, Wilson's 
lattice QCD has proved to be the best suited theoretical framework. Here we 
argue that the quantum link formulation of QCD is ideally suited for quantum 
simulations that address the real-time dynamics of quarks and gluons as well as 
QCD at high quark density.

In Sections 2 and 3, we will discuss quantum simulators for Abelian and
non-Abelian quantum link models, which share qualitative features with QCD.
The concluding Section 4 outlines what needs to be done in order to ultimately 
approach quantum simulations of QCD itself.

\section{Quantum Simulators for Abelian Gauge Theories}

Quantum simulators encode the state of the simulated quantum system as quantum
information in its discrete quantum states. Ultracold atomic gases in optical
lattices are used as quantum simulators for condensed matter systems including
bosonic or fermionic Hubbard models. The information about the corresponding
quantum states is then encoded in the discrete positions of fermionic or 
bosonic atoms in the wells of the optical lattice as well as in the discrete 
internal states of the atoms. The gauge theories of particle physics, including 
QCD, are usually formulated with continuous gauge field degrees of freedom.
Even in Wilson's lattice QCD, in which the gauge field resides on the links of a
discrete lattice, the link variables themselves are continuous variables which
give rise to an infinite-dimensional link Hilbert space. It is not clear
how Wilson's continuously varying link variables, which take values in the gauge
group ($SU(3)$ for QCD), could be represented with a handful of discrete states,
for example, of ultracold atoms. Quantum link models provide an alternative
lattice regularization of gauge field theories including QCD \cite{Bro99}, in 
which the link Hilbert space is finite-dimensional, and can thus be naturally 
represented, for example, with discrete states of ultracold atoms in an optical 
lattice.

In order to introduce quantum link models in a simple context, we first consider
a $U(1)$ gauge theory. Quantum link models share most of the structure of 
Wilson's lattice gauge theory in the Hamiltonian formulation, in which space is
replaced by a cubic lattice and time remains continuous. In this formulation,
the gauge field is represented by classical parallel transporter link variables
$U_{xy} =\exp(i \varphi_{xy}) \in U(1)$, which take values in the gauge group. 
Their canonically conjugate momenta are electric field operators 
$E_{xy} = - i \partial/\partial \varphi_{xy}$, which are again associated with 
the links, and which obey the commutation relations
\begin{equation}
\label{commuteEU}
[E_{xy},U_{x'y'}] = \delta_{xx'} \delta_{yy'} U_{xy}, \quad 
[E_{xy},U_{x'y'}^\dagger] = - \delta_{xx'} \delta_{yy'} U_{xy}^\dagger,
\end{equation}
such that operators associated with different links commute with each other.
Also introducing staggered fermion creation and annihilation operators 
$\psi_x^\dagger$, $\psi_x$, which obey canonical anti-commutation relations
\begin{equation}
\{\psi_x^\dagger,\psi_y\} = \delta_{xy}, \quad 
\{\psi_x^\dagger,\psi_y^\dagger\} = \{\psi_x,\psi_y\} = 0,
\end{equation}
the Hamiltonian of lattice QED takes the form
\begin{equation}
\label{Hamiltonian}
H_{\text{QED}} = - t \sum_{\langle x y \rangle} s_{xy} \left(\psi_x^\dagger U_{xy} \psi_y 
+ \psi_y^\dagger U_{xy}^\dagger \psi_x \right) + 
m \sum_x s_x \psi_x^\dagger \psi_x +
\frac{e^2}{2} \sum_{\langle x y \rangle} E_{xy}^2 - \frac{1}{4 e^2}
\sum_\Box \left(U_\Box + U_\Box^\dagger\right).
\end{equation}
Here $t$ is a hopping parameter, $m$ is the fermion mass, and $e$ is the 
electric charge. The factor $s_x = (-1)^{x_1 + \dots + x_d}$ is associated with the 
sites and the factor $s_{xy}$ is associated with the links $\langle x y \rangle$
connecting nearest-neighbor sites $x$ and $y$ of the $d$-dimensional spatial 
lattice. The links in the 1-direction have $s_{xy} = 1$, the ones in the 
2-direction have $s_{xy} = (-1)^{x_1}$, and those in the $k$-direction have 
$s_{xy} = (-1)^{x_1 + \dots + x_{k-1}}$. The product 
$U_\Box =  U_{wx} U_{xy} U_{yz} U_{zw}$ around an elementary lattice plaquette
$\Box$ represents the magnetic field energy. The Hamilton operator is gauge 
invariant, i.e.\ it commutes with the generators of local gauge transformations 
\begin{equation}
\label{Gauss}
G_x = \psi_x^\dagger \psi_x + 
\sum_k \left(E_{x,x+\hat k} - E_{x-\hat k,x}\right), \quad [H,G_x] = 0,
\end{equation}
where $\hat k$ is a unit-vector in the $k$-direction.
The gauge generators $G_x$ represent the difference of the local charge density
$\psi_x^\dagger \psi_x$ and the lattice divergence of the electric field. 
Physical states $|\Psi\rangle$ must obey the Gauss law $G_x |\Psi\rangle = 0$, 
i.e.\ they must be gauge invariant. Since $U_{xy}$ is a continuously varying 
link variable, $E_{xy} \in \Z$ can take infinitely many values, and thus the 
Hilbert space of Wilson's lattice gauge theory is infinite-dimensional, already 
for a single link.

Quantum link models realize continuous gauge symmetries with discrete gauge
variables and thus in a finite-dimensional link Hilbert space. In an Abelian 
$U(1)$ quantum link model, the link operator as well as the electric field 
operator are given by quantum spin operators $\vec S_{xy}$ associated with the
link $xy$
\begin{equation}
U_{xy} = S_{xy}^1 + i S_{xy}^2 = S_{xy}^+, \quad 
U_{xy}^\dagger = S_{xy}^1 - i S_{xy}^2 = S_{xy}^-, \quad 
E_{xy} = S_{xy}^3.
\end{equation}
The quantum link operators $U_{xy}$ and $U_{xy}^\dagger$ act as raising and 
lowering operators of electric flux. The eqs.(\ref{commuteEU}),
(\ref{Hamiltonian}), and (\ref{Gauss}) of the Wilson theory still hold unchanged
in quantum link models. The only difference, which allows the drastic 
simplification of a finite-dimensional Hilbert space per link, is that now
$[U_{xy},U_{x'y'}^\dagger] = 2 \delta_{xx'} \delta_{yy'} E_{xy} \neq 0$. 

Although the simple $(2+1)$-d $U(1)$ quantum link model with a quantum spin 
$\frac{1}{2}$ is not completely equivalent to the Wilson theory, it shares the
important feature of confinement. The dynamics of this model is illustrated in 
Fig.1. 
\begin{figure}[h]
\begin{center}
\includegraphics[width=0.65\textwidth]{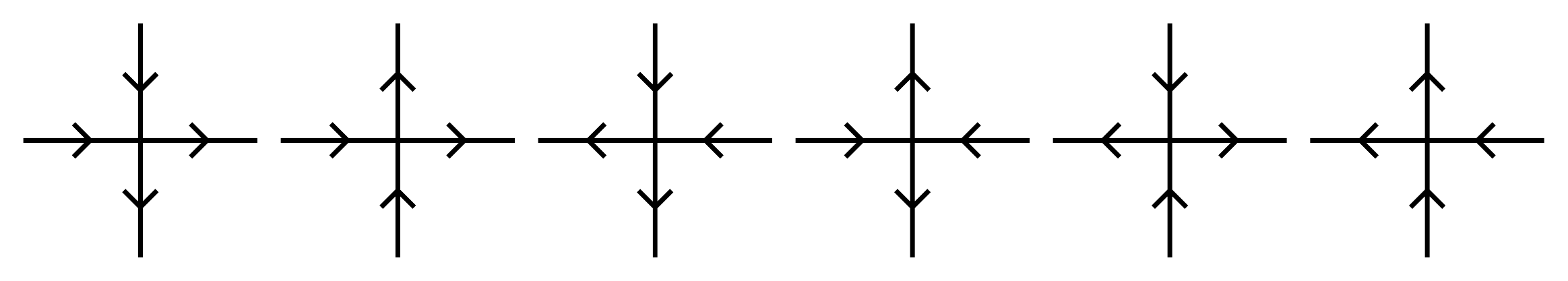} \\
\includegraphics[width=0.33\textwidth]{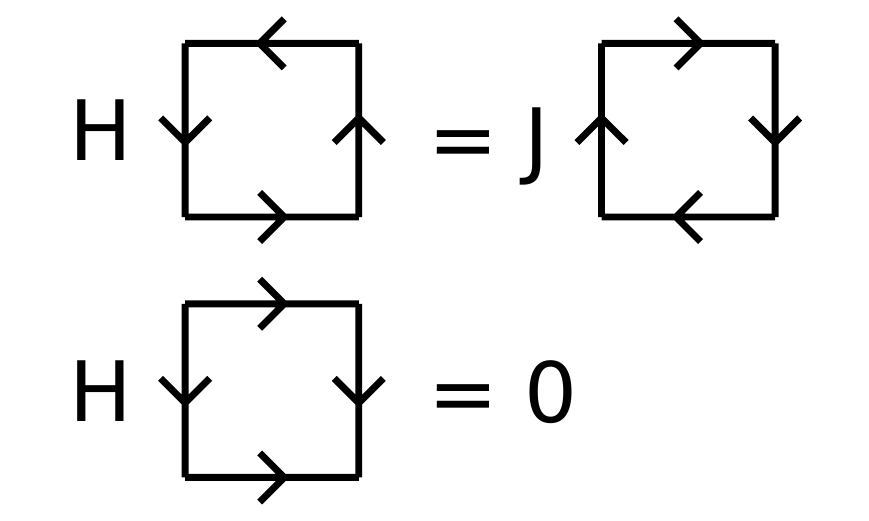}
\includegraphics[width=0.25\textwidth]{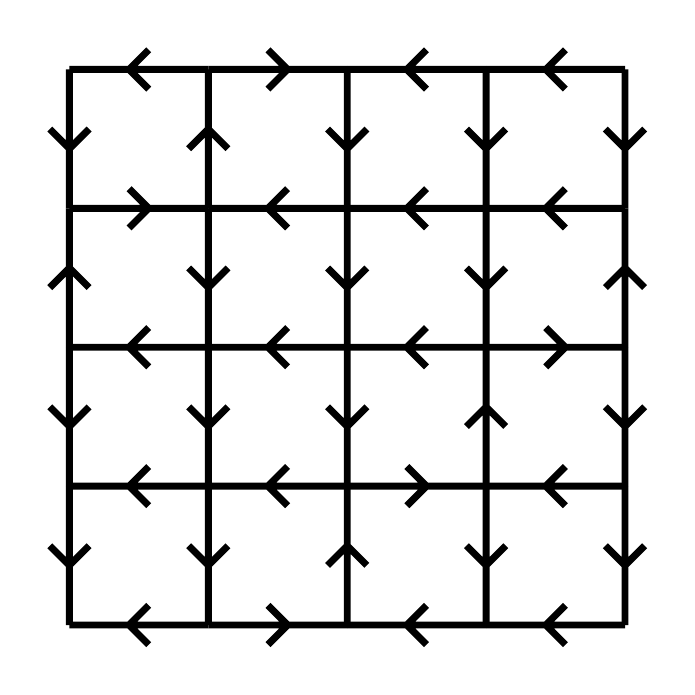}
\includegraphics[width=0.33\textwidth]{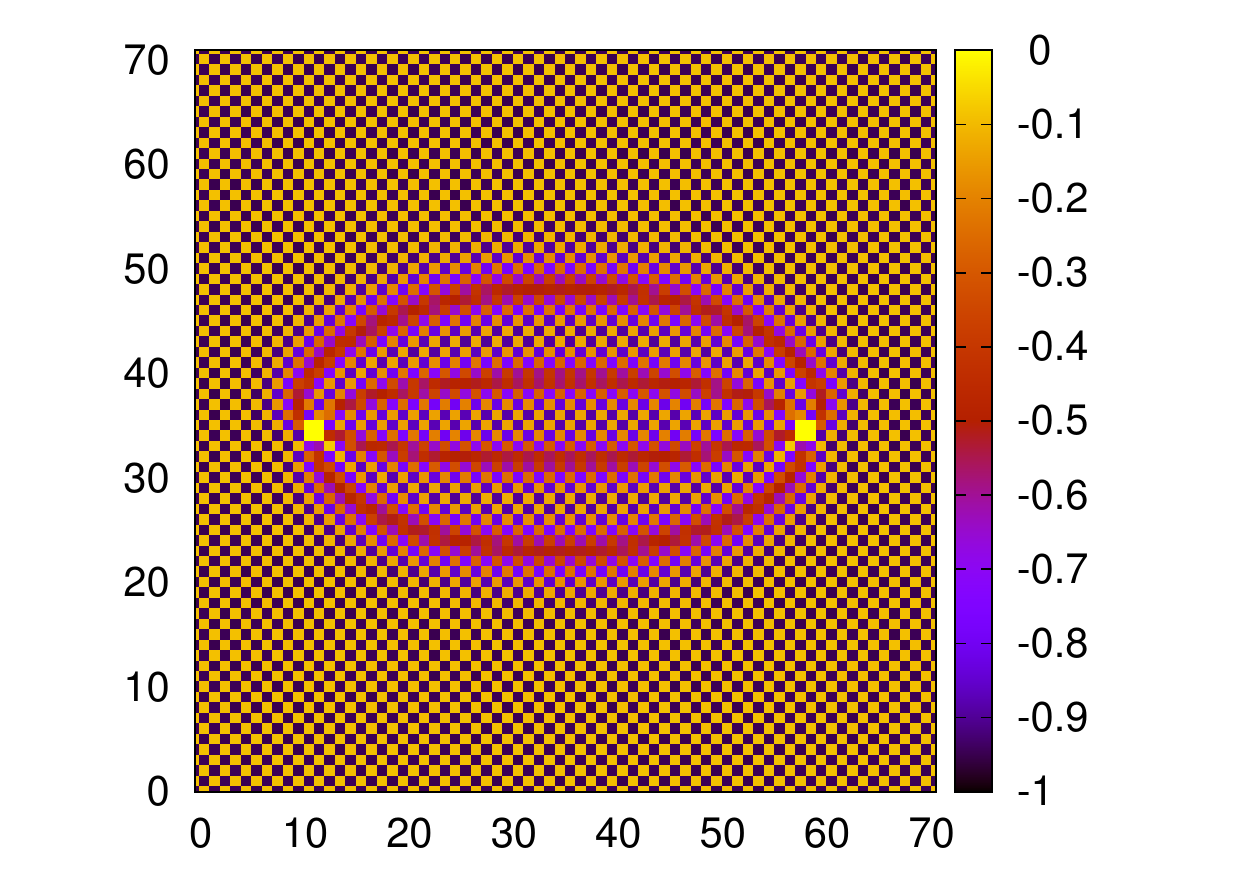}
\caption{\textit{Top: The six flux configurations that satisfy the Gauss law
in the $(2+1)$-d $U(1)$ quantum link model. Bottom left: The Hamiltonian 
reverses the direction of a closed loop of flux, and annihilates all other flux 
states. Bottom middle: A flux configuration on a $4 \times 4$ lattice with 
periodic boundary conditions, with the fluxes obeying Gauss' law at every site 
$x$. Bottom right: Energy density $- (U_\Box + U_\Box^\dagger)$ of the string 
connecting two external charges $Q = \pm 2$. The string splits into four 
strands, each carrying electric flux $\frac{1}{2}$ \cite{Ban13a}.}}
\end{center}
\end{figure}
Both digital and analog quantum simulator constructions have been proposed for 
this system. Highly excited Rydberg atoms in an optical lattice, which can be 
addressed individually with external lasers, have strong long-range 
dipole-dipole interactions, thus allowing the entanglement of a number of atoms 
with a single control atom. This forms the basis of digital quantum simulator 
constructions for $U(1)$ quantum link models \cite{Wei10,Tag13}. These use 
control atoms at lattice sites to ensure Gauss' law, and at plaquette centers 
to flip electric flux loops, while other Rydberg atoms act as qubits that
represent the quantum link variables. Some analog quantum simulator 
constructions again use ultracold atoms in optical lattices 
\cite{Bue05,Zoh11,Zoh12}, while others are based on superconducting circuits on 
a chip \cite{Mar13,Mar14}.

In \cite{Ban12,Zoh13,Zoh13b} Bose-Fermi mixtures were proposed to quantum 
simulate $U(1)$ quantum links coupled to dynamical staggered fermions. Then the 
string that connects external static charges can break by the creation of a 
dynamical charge-anti-charge pair. As illustrated in Fig.2, the proposed 
quantum simulator can investigate string-breaking in real time. 
\begin{figure}[h]
\begin{center}
\includegraphics[width=0.33\textwidth]{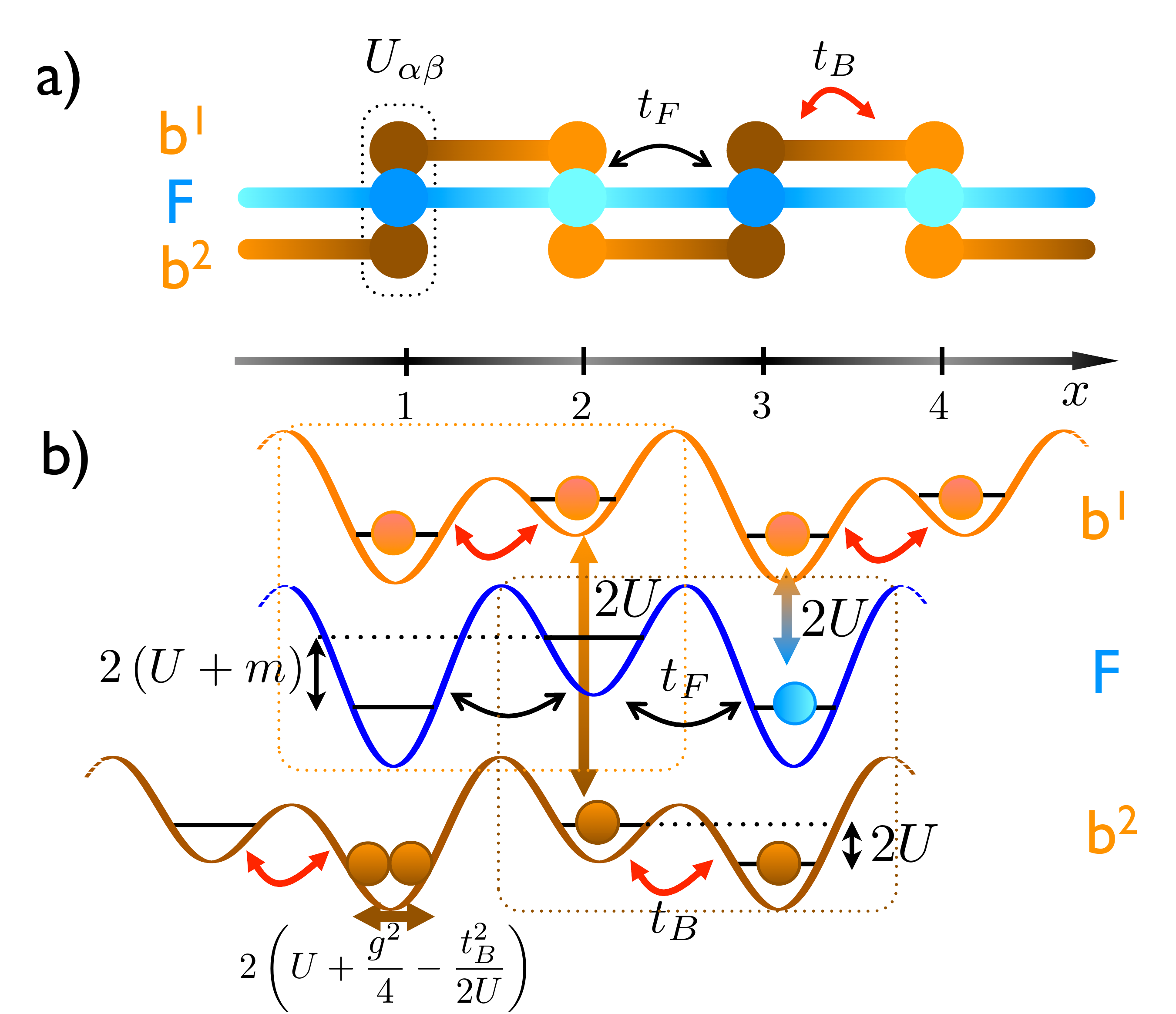} \hskip1cm
\includegraphics[width=0.37\textwidth]{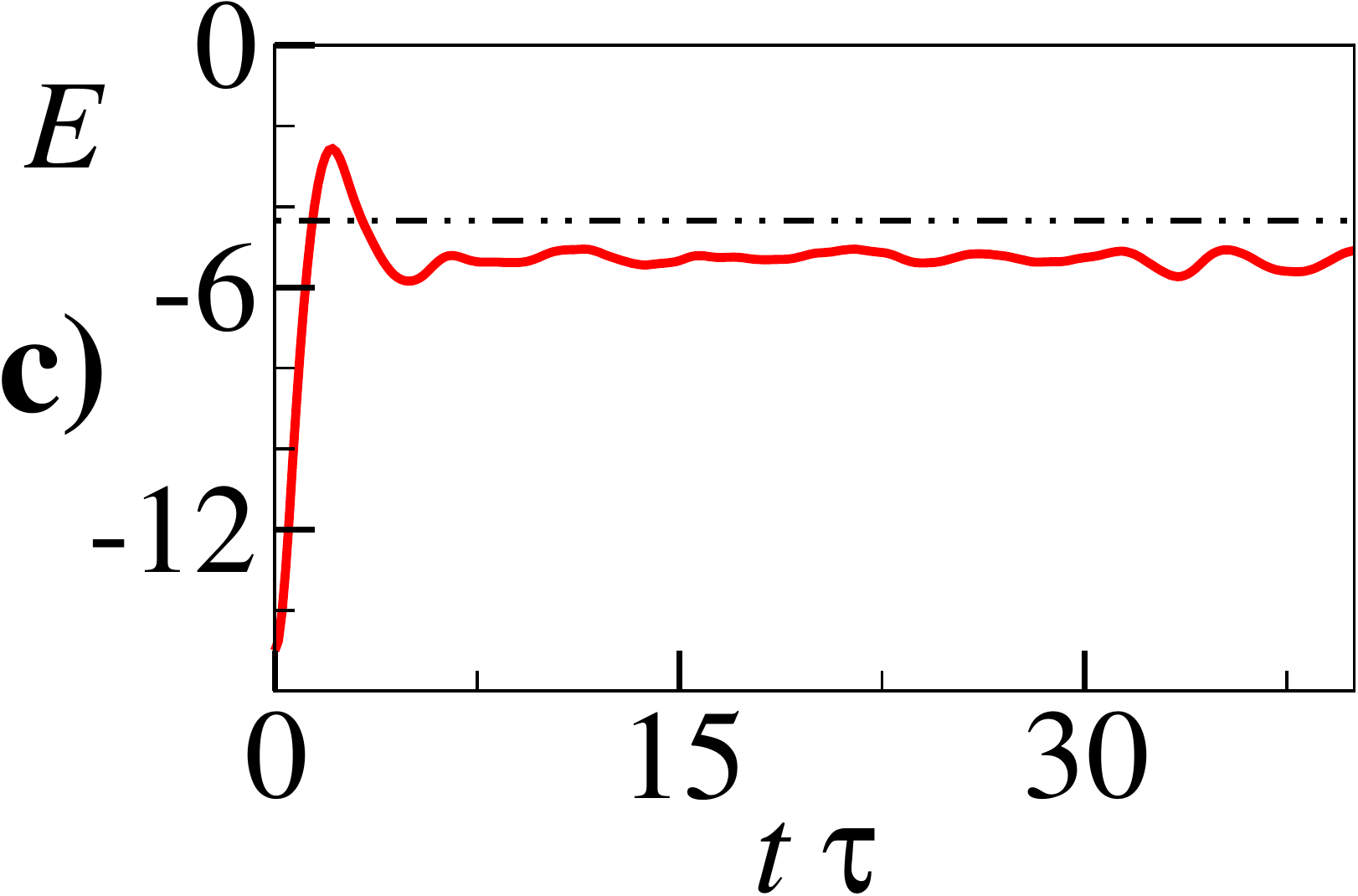}
\caption{\textit{a) Quantum simulator of a $(1+1)$-d $U(1)$ quantum link model 
with staggered fermion matter based on a Bose-Fermi mixture of ultracold atoms 
hopping in a 3-strand optical superlattice. b) Mass $m$, hopping $t$, and 
interaction $U$, $g$ parameters are determined by the shape of the optical 
superlattice. Energy conservation favors the simultaneous hopping of fermions 
and bosons, thus ensuring gauge invariance. c) Dynamical string breaking in 
real time: the string energy is converted into the mass of a charge-anti-charge
pair, thus depleting the electric flux $E$ of the string \cite{Ban12}.}}
\end{center}
\end{figure}

\section{Quantum Simulators for non-Abelian Gauge Theories}

Non-Abelian gauge theories can also be regularized with quantum link models.
Like a Wilson-type $SU(N)$ link variable, a quantum link in an $SU(N)$ gauge
theory still is an $N \times N$ matrix. However, its matrix elements 
$U_{xy}^{ij}$ are no longer complex numbers, but non-commuting operators. 
Together with non-Abelian electric flux operators $L^a_{xy}$ and $R^a_{xy}$ 
residing on the left and right end of a link, and an Abelian electric flux
$E_{xy}$, the quantum link operators form the embedding algebra $SU(2N)$. The
$2 N^2$ real and imaginary parts of $U_{xy}^{ij}$, together with the $2(N^2-1)$
$SU(N)$ generators $L^a_{xy}$ and $R^a_{xy}$ and the generator $E_{xy}$, indeed
provide the $2N^2 + 2(N^2-1) + 1 = 4N^2-1$ generators of $SU(2N)$. Quantum 
link models can realize exact continuous $SU(N)$ gauge symmetry with any 
finite-dimensional representation of $SU(2N)$, while the Wilson theory again
has an infinite-dimensional Hilbert space per link. In particular, $SU(3)$ 
quantum link QCD uses a 20-dimensional representation of the embedding algebra
$SU(6)$. Just as in the standard Wilson formulation, in the quantum link
regularization of QCD the Hamiltonian is given by
\begin{equation}
\label{HamiltonianQCD}
H_{\text{QCD}} = 
- t \sum_{\langle x y \rangle} s_{xy} \left(\psi_x^\dagger U_{xy} \psi_y + 
\psi_y^\dagger U_{xy}^\dagger \psi_x \right) + 
m \sum_x s_x \psi_x^\dagger \psi_x +
\frac{g^2}{2} \sum_{\langle x y \rangle} (L_{xy}^2 + R_{xy}^2) - \frac{1}{4 g^2}
\sum_\Box \mbox{Tr} \left(U_\Box + U_\Box^\dagger\right),
\end{equation}
which commutes with the infinitesimal generators of gauge transformations
\begin{equation}
\label{generatorQCD}
G_x^a = \psi_x^{i \dagger} \lambda^a_{ij} \psi_x^j + 
\sum_k \left(L_{x,x+\hat k} + R_{x-\hat k,x}\right), \quad
[G^a_x,G^b_y] = 2 i \delta_{xy} f_{abc} G^c_x.
\end{equation}
It is a remarkable feature of quantum link models, that the gauge degrees of 
freedom can be expressed as fermion bilinears
\begin{equation}
L^a_{xy} = c^{i \dagger}_{x,+} \lambda^a_{ij} c^j_{x,+}, \quad
R^a_{xy} = c^{i \dagger}_{y,-} \lambda^a_{ij} c^j_{y,-}, \quad
E_{xy} = \frac{1}{2}(c^{i \dagger}_{y,-} c^i_{y,-} - c^{i \dagger}_{x,+} c^i_{x,+}), 
\quad U^{ij}_{xy} = c^i_{x,+} c^{j \dagger}_{y,-}.
\end{equation}
The fermionic constituents of the gauge field are called rishons and obey
standard anti-commutation relations. The rishon dynamics is illustrated in 
Fig.3. 
\begin{figure}[h]
\begin{center}
\includegraphics[width=0.33\textwidth]{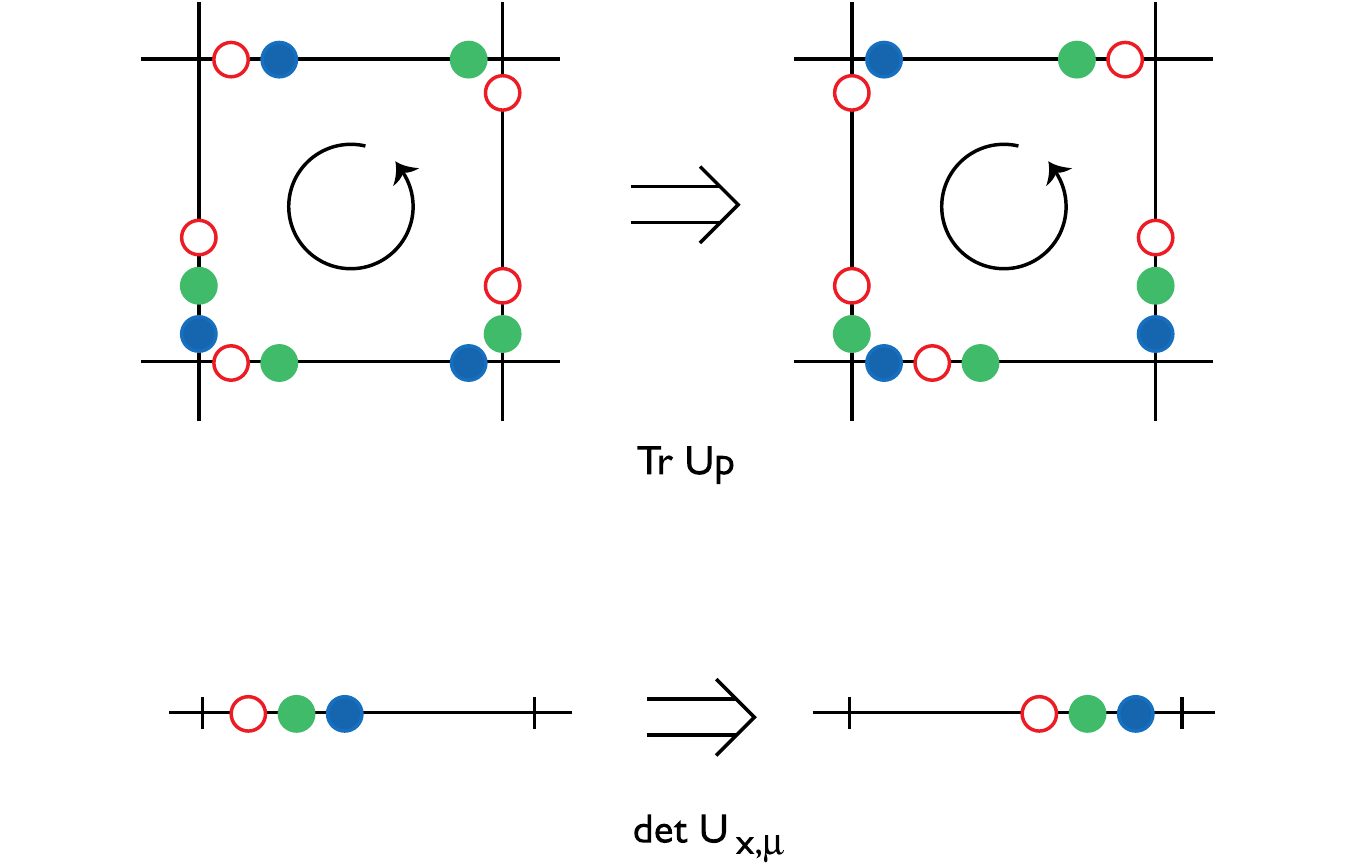}
\includegraphics[width=0.25\textwidth]{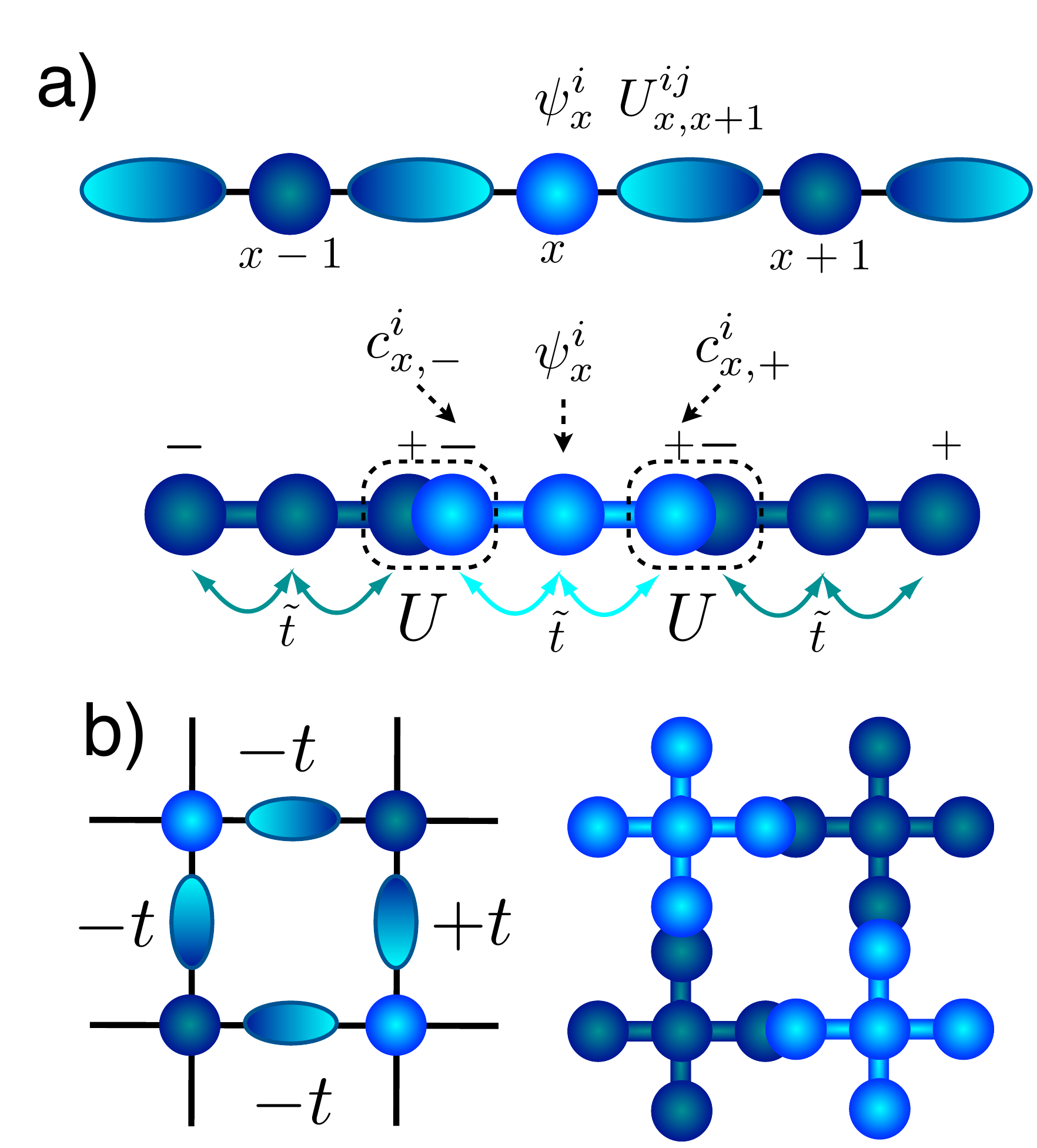}
\includegraphics[width=0.37\textwidth]{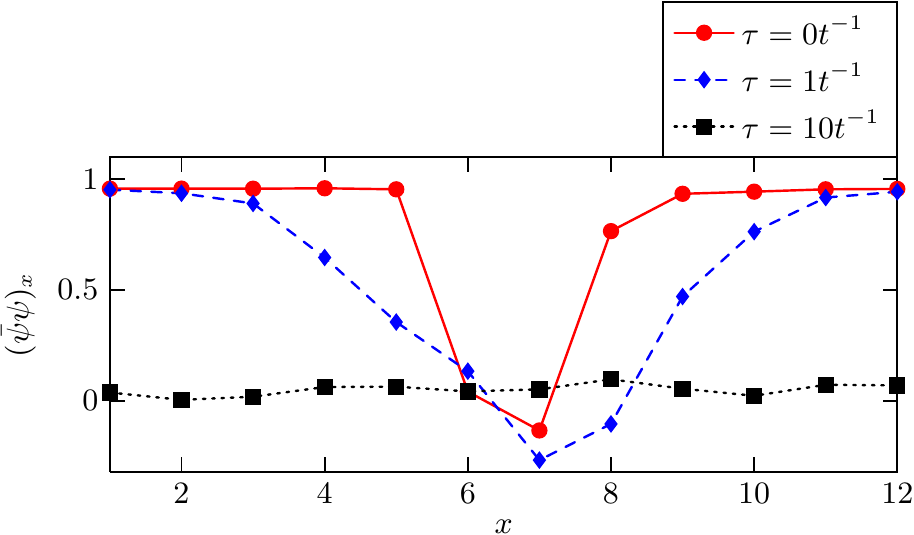}
\caption{\textit{Left: In an $SU(3)$ quantum link model, three fermionic rishons
of different colors reside on each link. The Hamiltonian moves the rishons 
around plaquettes or along links, like the beads of an abacus. Middle: Quantum 
simulator for an $SU(N)$ gauge theory with ultracold alkaline-earth atoms 
representing ``quarks'' or rishon constituents of ``gluons'' in a 1-d (a) or 
2-d (b) optical superlattice. Depending on its position in the optical lattice, 
an alkaline-earth atom either embodies a ``quark'' or a rishon. Gauge invariance
is protected by the internal $SU(2I+1)$ symmetry of the nuclear spin $I$ of the 
atoms. Right: Real-time $\tau$ evolution of the chiral order parameter profile 
$(\overline \psi \psi)_x$ in a $U(2)$ gauge theory, mimicking the expansion of 
a hot ``quark-gluon'' plasma \cite{Ban13}.}}
\end{center}
\end{figure}

A digital quantum simulator construction for an $SU(2)$ pure gauge theory using 
Rydberg atoms in an optical lattice has been proposed in \cite{Tag13}. An analog
quantum simulator for an $SU(2)$ gauge theory coupled to fermionic matter using 
a Fermi-Bose mixture of atoms in an optical superlattice has been constructed
theoretically in \cite{Zoh13a,Zoh13b}. An alternative analog proposal embodies
the rishons of quantum link models with fermionic  $^{87}$Sr or $^{173}$Yb 
alkaline-earth atoms moving in an optical superlattice (c.f.\ Fig.3) 
\cite{Ban13}. The interactions between the atoms can be engineered using 
Feshbach resonances. The $SU(N)$ color of a rishon is encoded in the Zeeman 
states of the nuclear spin $I$, which have a remarkably precise $SU(2I+1)$ 
symmetry that naturally protects $SU(N)$ gauge invariance. Each link has two 
rishon-sites, one at each end. In addition, there are the ordinary lattice sites
where ``quarks'' reside. When an alkaline-earth atom sits on a rishon-site it 
embodies a fermionic constituent of a ``gluon''. The same atom represents a 
``quark'' when it hops to an adjacent ordinary lattice site. Even a simple 
$U(2)$ toy-model gauge theory, which can be realized with a single rishon per
link, is quite interesting, because it has a spontaneously broken chiral 
symmetry, whose dynamics can be investigated in real time (c.f.\ Fig.3). 

The simple $U(2)$ toy-model gauge theory does not have ``baryons'', because 
baryon number is part of the gauge symmetry. Another toy-model gauge theory 
that does contain ``baryons'' and thus can be used to mimic qualitative 
features of nuclear physics has an $SO(3)$ gauge symmetry. When one puts 
``quarks'' in the adjoint triplet representation of $SO(3)$, such that they have
the same color quantum numbers as $SO(3)$ ``gluons'', one can form color-singlet
``baryons'' that consist of a single ``quark'' confined to a ``gluon''. Such a 
system can be quantum simulated using magnetic atoms with dipole-dipole
interactions in an optical lattice \cite{Dal14}. Even a simple $(1+1)$-d 
toy-model $SO(3)$ gauge theory has a $\Z(2)$ chiral symmetry which is 
spontaneously broken in the vacuum, but restored at finite baryon density.
Fig.4 shows the energy difference $\Delta E$ between the vacuum state and the 
first excited state as a function of the spatial volume $L$. At zero baryon 
density $n_B$, $\Delta E$ decreases exponentially with $L$, thus indicating 
vacuum degeneracy, and hence spontaneous $\Z(2)$ chiral symmetry breaking, in
the infinite-volume limit. At higher baryon densities, $n_B \geq \frac{1}{2}$,
$\Delta E$ no longer decreases exponentially, thus indicating that chiral
symmetry gets restored. A quantum simulator could be used to investigate the
high baryon-density phase diagram of this toy-model, which may even show
color-superconductivity.
\begin{figure}[h]
\begin{center}
\includegraphics[width=0.4\textwidth]{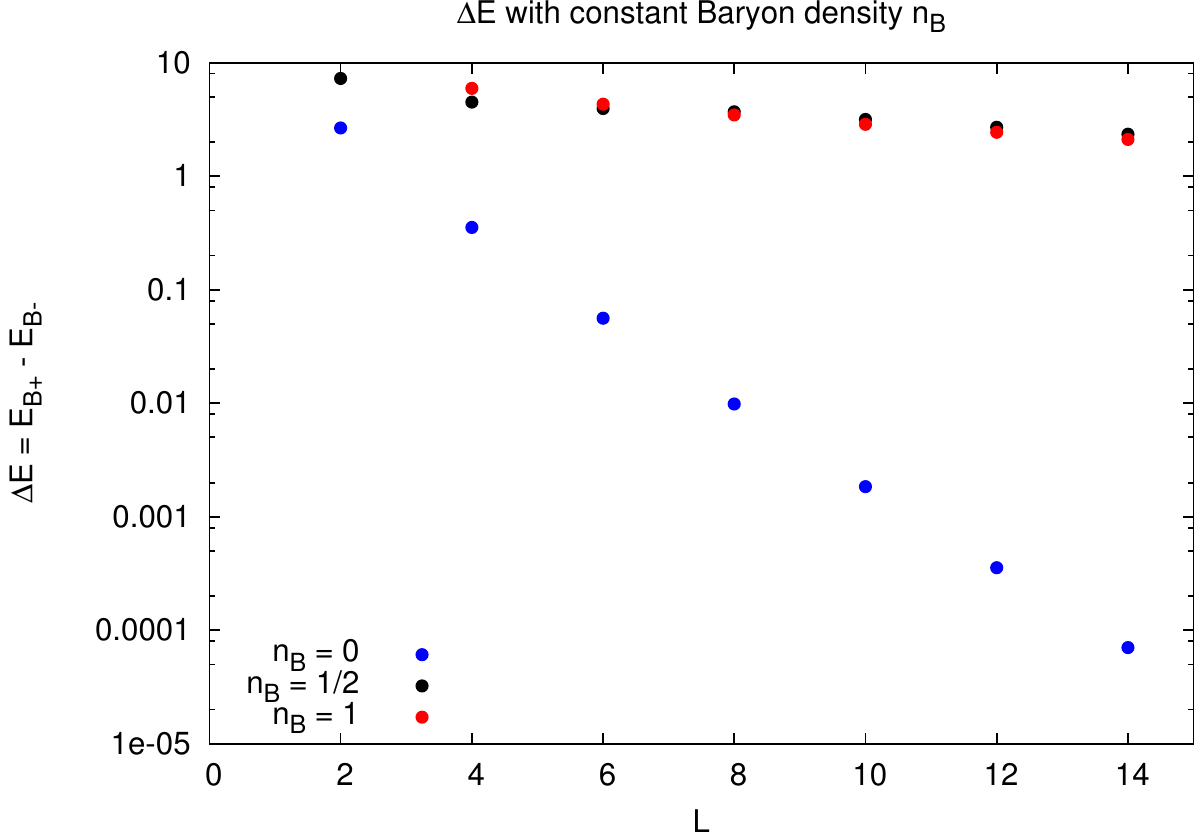} \hskip0.1cm
\includegraphics[width=0.08\textwidth]{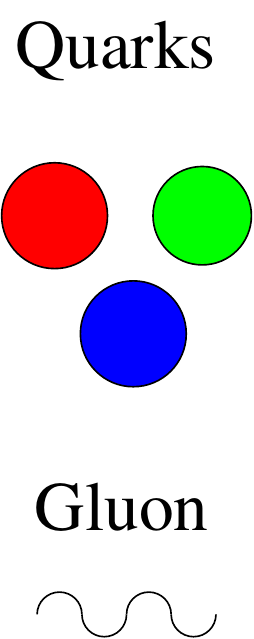}
\includegraphics[width=0.17\textwidth]{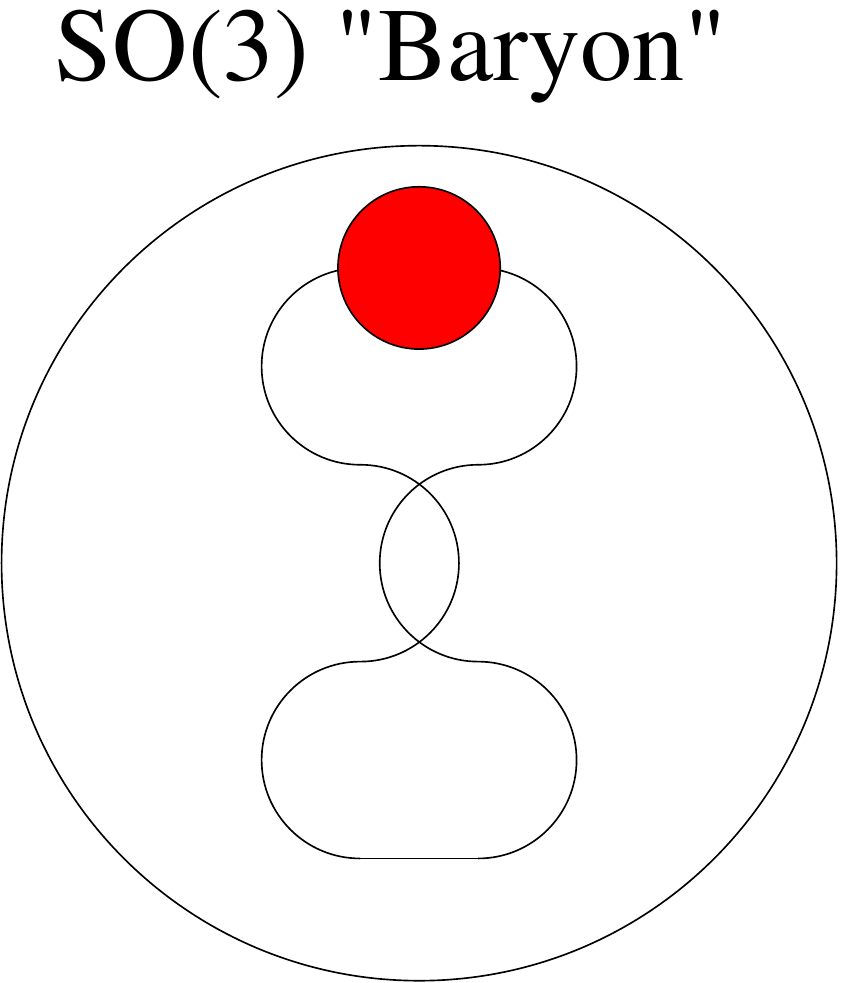}
\includegraphics[width=0.2\textwidth]{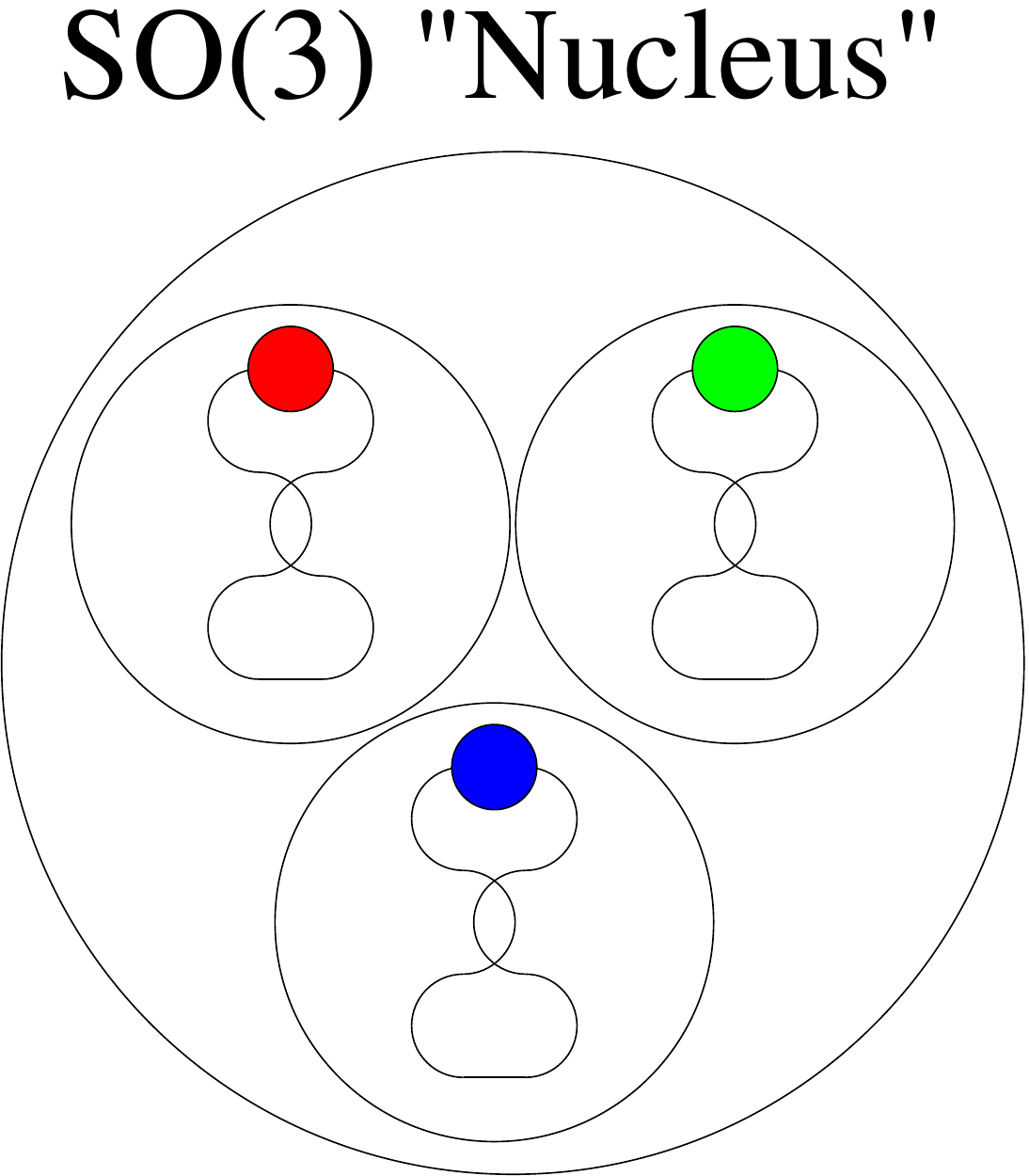}
\caption{\textit{Left: Chiral symmetry restoration in a $(1+1)$-d $SO(3)$ 
toy-model gauge theory with adjoint fermions. Right: Adjoint triplet ``quarks'' 
and ``gluons'' in a toy-model $SO(3)$ gauge theory. An $SO(3)$ ``baryon'' 
consists of a single triplet ``quark'' and a triplet ``gluon'' forming a color 
singlet. Several baryons may form an $SO(3)$ ``nucleus''.}}
\end{center}
\end{figure}

\section{Conclusions}

Quantum link models provide an alternative non-perturbative regularization of
lattice gauge theories, which is ideally suited for quantum simulation, for
example, using ultracold atomic gases in optical lattices. This holds the 
promise to address very challenging problems, such as the real-time and high
density dynamics of strongly interacting matter, which, due to very severe sign
or complex action problems, cannot be addressed with classical computers. While 
it is hard to predict when quantum simulations of full QCD might become 
feasible, toy-model gauge theories that share qualitative features with QCD 
should be possible to quantum simulate within the next few years. In this way 
one can gain qualitatively new insights into highly non-trivial dynamics of 
Abelian and non-Abelian gauge theories, such as string breaking in real time, 
the expansion of a chirally restored ``fireball'', or the restoration of chiral
symmetry at high ``baryon'' density. If such quantum simulations can be
performed successfully, in the long run quantum simulators may advance to a 
more and more accepted tool of theoretical physics. Before universal quantum
computers become available, one may thus hope to eventually develop a 
special purpose quantum simulator to address currently unsolvable 
problems in QCD. The quantum link formulation of QCD provides a theoretical
framework as well as a
vision how the path towards this long-term goal could look like. Along this 
path, one must overcome severe challenges: proceeding to higher dimensions, 
realizing the full chiral symmetry of light quarks and the correct gauge group 
$SU(3)$, and, finally, taking the continuum limit. None of these difficult tasks
seem insurmountable, at least in principle. While the ultimate goal of this
research may be in the distant future, starting to work in this direction seems
timely now. Interdisciplinary research between quantum optics, atomic, and
particle physics, along the lines suggested here, holds the promise to deepen
our understanding of the fundamental gauge structures underlying all of physics.
There is a lot of exciting physics to be explored along the way towards a 
full QCD quantum simulator as a reliable tool for nuclear and particle physics.

\section*{Acknowledgments}

I like to thank the organizers of Quark Matter 2014 for giving me the
opportunity to present this work at the conference.
Several results discussed here were obtained together with D.\ Banerjee, 
M.\ B\"ogli,  M.\ Dalmonte, F.-J.\ Jiang, M.\ M\"uller, E.\ Rico Ortega, 
P.\ Stebler, P.\ Widmer, and P.\ Zoller. I thank them for a very pleasant and 
fruitful collaboration. The research leading to these results has received 
funding from the Schweizerischer Na\-tio\-nal\-fonds and from the European 
Research Council under the European Union's Seventh Framework Programme 
(FP7/2007-2013) ERC grant agreement 339220.

\end{document}